\documentclass{PoS}
\usepackage{wrapfig}
\def\GeV{{\rm GeV}}
\def\TeV{{\rm TeV}}

\title{Updates of PDFs in the MSTW framework}

\ShortTitle{Updates of PDFs in the MSTW framework}

\author{\speaker{R.S.~Thorne} \\
        Department of Physics and Astronomy, \\  
        University College London, WC1E 6BT, UK\\
        E-mail: \email{robert.thorne@ucl.ac.uk}}

\author{L.A.~Harland-Lang\\
        Institute for Particle Physics Phenomenology,\\
        University of Durham, DH1 3LE, UK \\ 
        E-mail: \email{lucian.harland-lang@durham.ac.uk}}

\author{A.D.~Martin\\
        Institute for Particle Physics Phenomenology,\\
        University of Durham, DH1 3LE, UK \\ 
        E-mail: \email{A.D.Martin@durham.ac.uk}}

\author{P.~Motylinski \\
        Department of Physics and Astronomy, \\  
        University College London, WC1E 6BT, UK\\
        E-mail: \email{p.motylinski@ucl.ac.uk}}

\abstract{I present results on updates on PDFs which are obtained within
the general framework which led to the MSTW2008 PDF sets. 
There are some theory and procedural improvements and a variety of
new data sets, including many relevant up-to-date LHC data. 
A new set of PDFs is very close to being finalised, with no significant 
changes expected to the preliminary PDFs shown here.}

\FullConference{XXII. International Workshop on Deep-Inelastic Scattering and Related Subjects \\
		 28 April - 2 May 2014 \\
		 Warsaw, Poland}

\begin{document}

\vspace{-0.0cm}

I present an update of the parton distribution functions (PDFs)  
presented by 
the MSTW collaboration \cite{Martin:2009iq}. The overall procedure for
obtaining these PDFs is in most ways very similar to this previous analysis.
However, there are a number of changes in theoretical treatment or 
procedures, and a large number of new or updated data sets included,
particularly from the LHC.

In the new analysis we continue to use the extended parameterisation with 
Chebyshev polynomials, and the additional freedom in deuteron nuclear 
corrections introduced in \cite{Martin:2012da} which led to a change in 
the $u_V-d_V$ distribution. We now also use the optimal GM-VFNS choice 
\cite{Thorne:2012az} which is smoother near
to the heavy flavour transition points, particularly at NLO.
We correct the $\nu + N \to \mu^+\mu^-$ cross-sections 
\cite{Goncharov:2001qe} for a missing small 
contribution, though checks show this has a very small effect on the 
strange quark distribution. However, also relevant for these data,
we have changed the 
value of the charm branching ratio to muons used, and 
additionally we now apply 
an uncertainty on the branching ratio which feeds into the  
PDFs. Specifically we use $B_{\mu} = 0.092 \pm 10\%$ 
from \cite{Bolton:1997pq}, which is not reliant on a PDF input
which might bias the result.
We update to a more recent determination of 
nuclear target corrections \cite{deFlorian:2011fp}. These improves 
the global fit quality by $\sim 25$ units, mainly in nuclear target 
structure functions. We now treat correlated systematic errors 
as being multiplicative not additive. Explicitly we use 
using the $\chi^2$ definition 
\begin{equation}
\chi^2=\sum_{i=1}^{N_{pts}}\left(\frac{D_i+\sum_{k=1}^{N_{corr}}r_k\sigma_{k,i}^{corr}-T_i}{\sigma_i^{uncorr}}\right)^2+\sum_{k=1}^{N_{corr}}r_k^2,
\end{equation}
where $\sigma_{k,i}^{corr}= \beta_{k,i}^{corr}T_i$
and $\beta_{k,i}^{corr}$ are the percentage errors. The additive 
definition, which previously we used for all but the normalisation 
uncertainty, would use $\sigma_{k,i}^{corr}= \beta_{k,i}^{corr}D_i$. 
Effectively if 
\begin{equation}
D_i + \sum_{k=1}^{N_{corr}}\beta_{k,i}^{corr}D_i \sim
f*D_i \qquad {\rm or} \qquad T_i - \sum_{k=1}^{N_{corr}}\beta_{k,i}^{corr}T_i \sim T_i/f,
\end{equation}
\begin{equation}
\hspace{-1cm}{\rm then} \qquad 
\chi^2\sim \left(\frac{f*D_i-T_i}{\sigma_i^{uncorr}}\right)^2\qquad
{\rm or} \qquad 
\chi^2\sim \left(\frac{D_i-T_i/f}{\sigma_i^{uncorr}}\right)^2 = 
\left(\frac{f*D_i-T_i}{f*\sigma_i^{uncorr}}\right)^2, 
\end{equation}
so with our new choice the uncorrelated errors scale with the data.
We make some other additional minor changes, but none 
have any significant impact.

\begin{figure}[htb]
\vspace{-1.5cm}
\centerline{\includegraphics[width=0.67\textwidth]{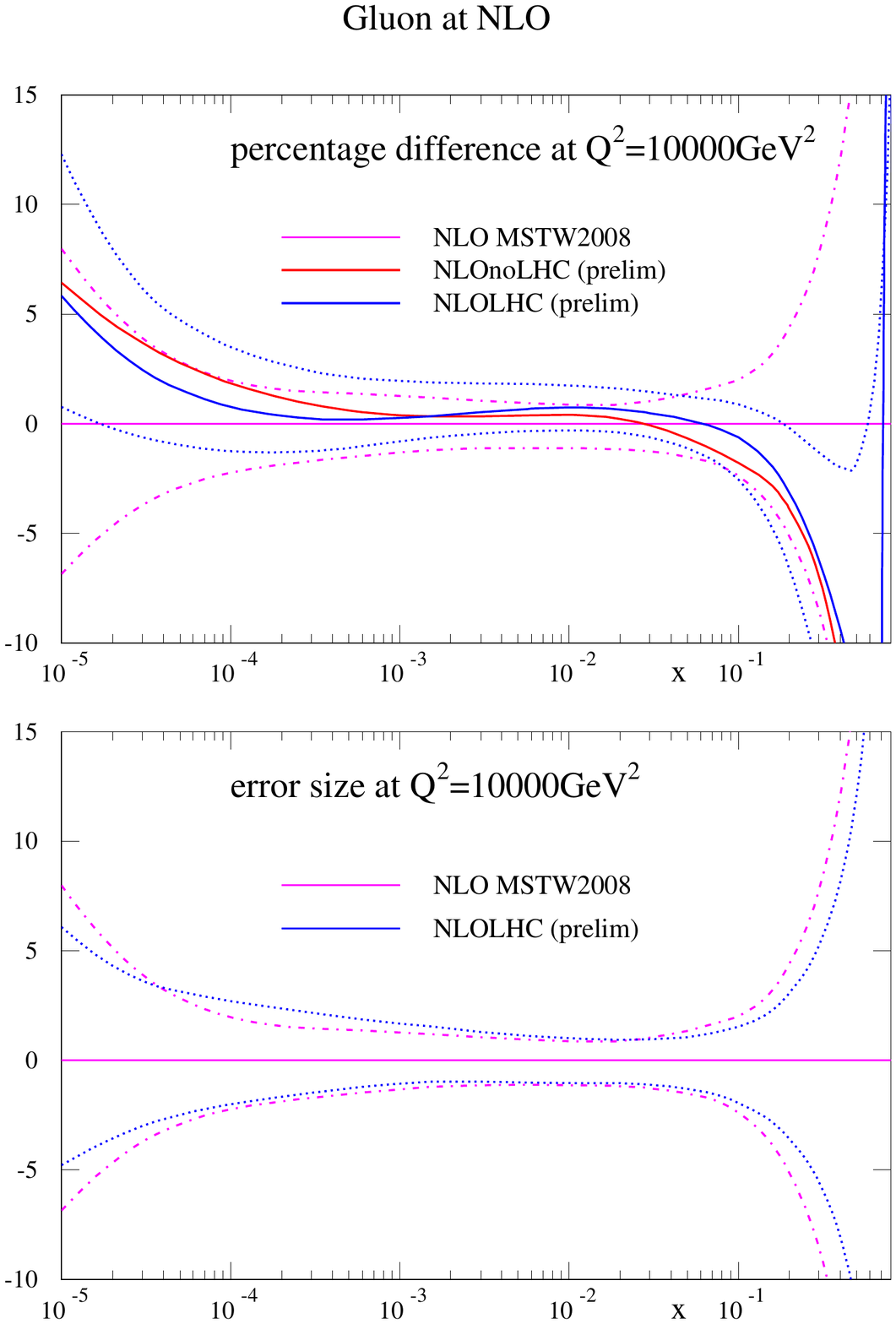}
\hspace{-2.6cm}\includegraphics[width=0.67\textwidth]{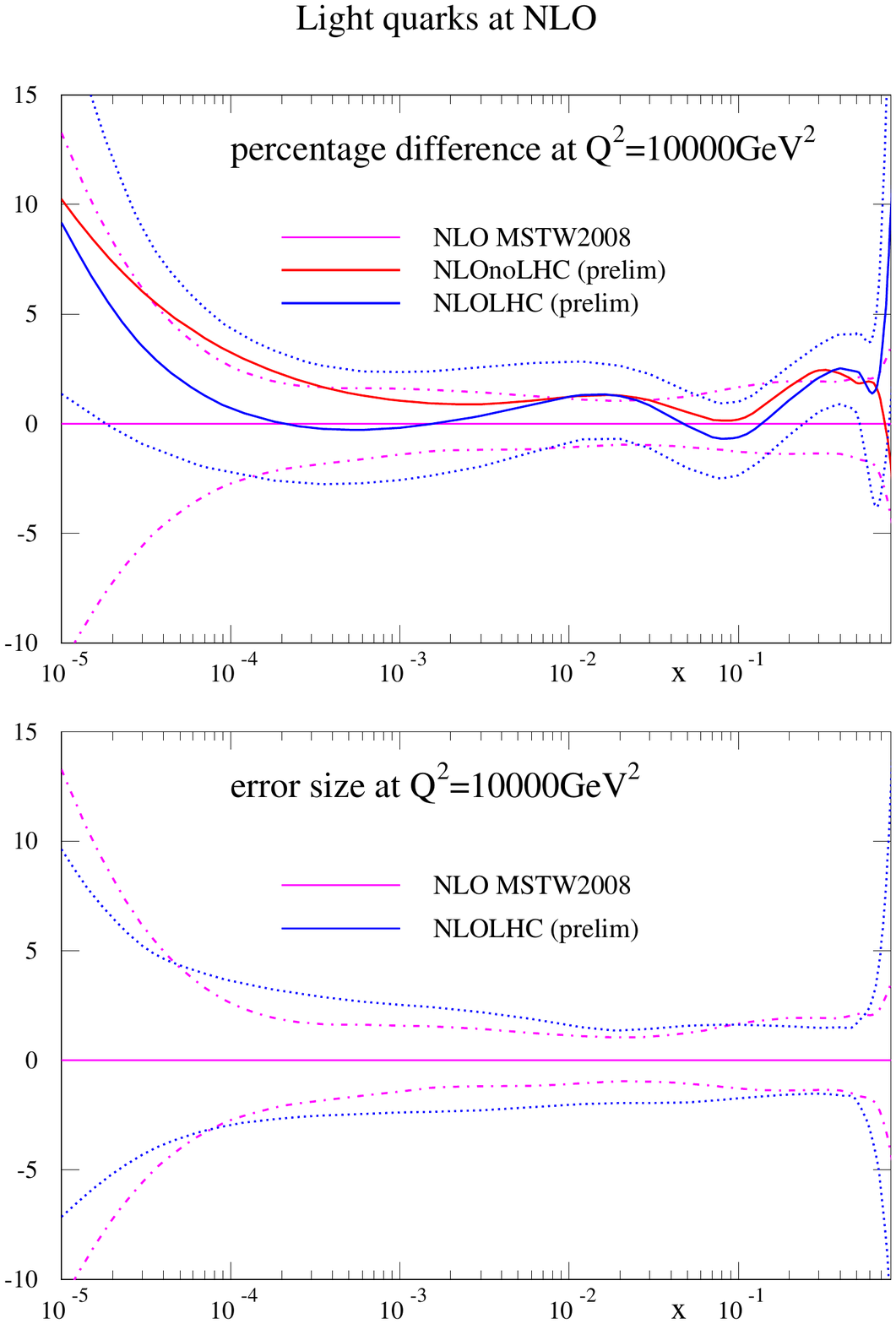}}
\vspace{-2.cm}
\caption{Comparison of the gluon and light quark distribution at NLO. }
\vspace{-0.5cm}
\label{Fig1} 
\end{figure}

\begin{figure}[htb]
\vspace{-1.5cm}
\centerline{\includegraphics[width=0.67\textwidth]{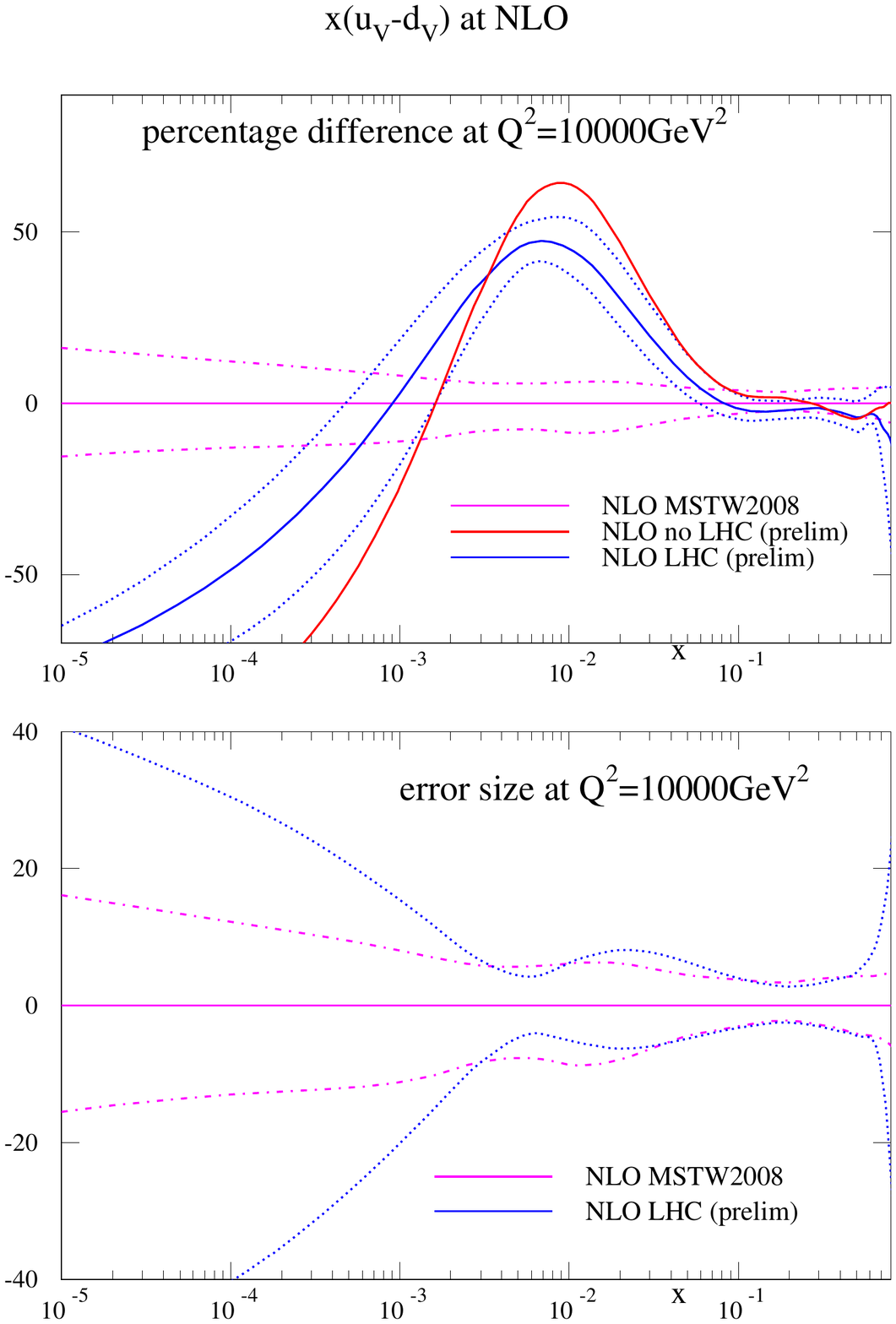}
\hspace{-2.6cm}\includegraphics[width=0.67\textwidth]{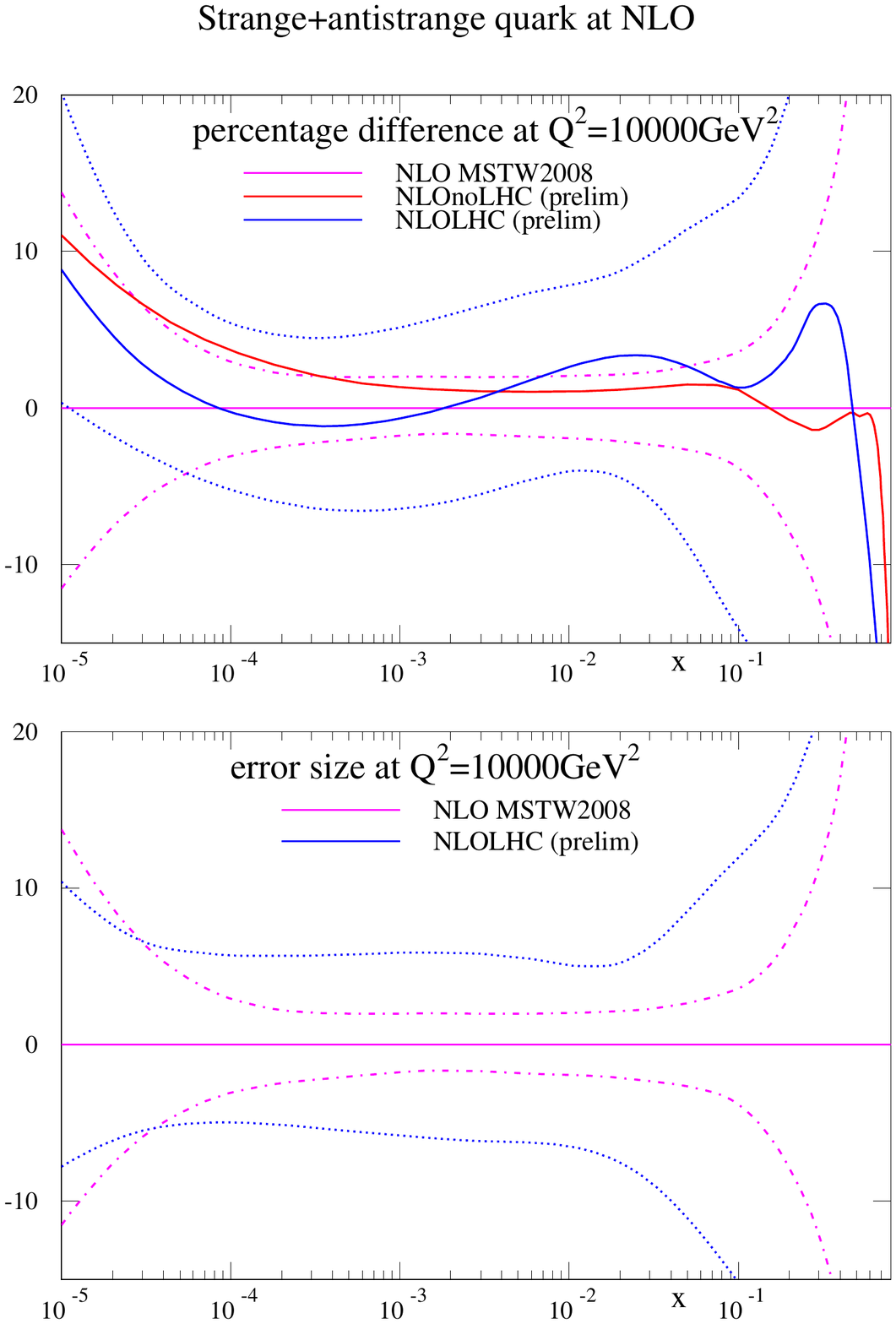}}
\vspace{-2.cm}
\caption{Comparison of the $u_V-d_V$ and $s+\bar s$ distribution at NLO. }
\vspace{-0.5cm}
\label{Fig2} 
\end{figure}

There are also various changes in non-LHC data sets. Most important is the 
replacement of HERA run I neutral and charged current data 
from H1 and ZEUS with the combined data set with the full treatment
of correlated errors \cite{Aaron:2009aa}. The fit to 
the data is very good and is slightly better at NNLO
than at NLO. We include the  HERA combined data on $F_2^c(x,Q^2)$
\cite{Abramowicz:1900rp}. Again the fit quality is about $\chi^2=1$ per
point.
%Inclusion of all direct published HERA $F_L(x,Q^2)$
%measurements. Undershoot data a little at lower $Q^2$, but $\chi^2$ 
%not much more than one per point.  
There is no inclusion of separate run II H1 and ZEUS data sets 
yet, since we wait instead for the Run II combination data. 
We include some updated Tevatron data sets, i.e. 
the CDF $W$-asymmetry data \cite{Aaltonen:2009ta}, the 
D0 electron asymmetry data \cite{Abazov:2008qv}
with $p_T> 25~\GeV$ based on 0.75 fb$^{-1}$
and new D0 muon asymmetry data \cite{Abazov:2013rja} for 
$p_T> 25 \GeV$ based on 7.3 fb$^{-1}$. 
We also include final numbers for CDF $Z$-rapidity data 
\cite{Aaltonen:2010zza} --
preliminary numbers were used in the MSTW2008 fit -- though this 
leads to very little change in PDFs. Overall the inclusion of the 
new HERA and Tevatron data and the change in procedures results in 
only a small change in the PDFs, other than already seen in 
$u_V-d_V$ \cite{Martin:2012da}, or in the best fit value of 
$\alpha_S(M_Z^2)$. At NLO $\alpha_S(M_Z^2) \to 0.1199$ from 
$0.1202$ and at NNLO $\alpha_S(M_Z^2)  \to 0.1180$  from
$0.1171$. The central value of the PDFs obtained from these changes 
are shown at NLO as a ratio to MSTW2008 in Figs. \ref{Fig1} and 
\ref{Fig2}.

As well as the updates mentioned above we now include a large variety of 
LHC data in the fits. A large component of this is 
rapidity-dependent vector boson production: ATLAS $W,Z$ cross sections
differential in rapidity \cite{Aad:2011dm}; CMS data on the asymmetry 
of $W$ bosons decaying to leptons \cite{Chatrchyan:2011jz,Chatrchyan:2012xt},
and $Z$ rapidity data \cite{Chatrchyan:2011wt}; 
LHCb data on $W$ and $Z$ rapidity distributions 
\cite{Aaij:2012vn,Aaij:2012mda}; and ATLAS high mass Drell Yan data
\cite{Aad:2013iua}. (We are yet to finalise fits to the data in
\cite{Chatrchyan:2013tia}.) 
The MSTW2008 data were known to describe the 
CMS $W$ asymmetry data and implicitly the asymmetry inherent in the ATLAS
$W$ data badly, but this was automatically improved 
enormously by the change in 
the small $x$ valence quarks in the MSTWCPdeut sets \cite{Martin:2012da}.
The $W^+/W^-$ asymmetry is no longer a problem at all 
for either the ATLAS and CMS data, with predictions and 
fit good, though slightly better at NLO. The fit quality for the ATLAS
$W,Z$ rapidity data before LHC data are included is $\chi^2\sim 1.6$ per 
point at NLO and $\chi^2\sim 2$ per point at NNLO. 
Inclusion of LHC data in the fit leads to some extra improvement at NLO, 
$\chi^2\sim 1.3$, with the strongest pull on the gluon PDF.
The quality also improves to $\chi^2\sim 1.3$ at NNLO, where the most 
obvious change is in the strange quark.
The fit quality to all other vector boson production data is good.  
We also include data on $\sigma_{t\bar t}$ from 
the combined cross section measurement from D0 and CDF \cite{Aaltonen:2013wca},
and all published data from ATLAS and CMS, using $m_{t}^{\rm pole} = 172.5~\GeV$ 
(the value used in the Tevatron combination) with an error of $1~\GeV$. 
Predictions and fit results are good, with  NLO preferring masses
slightly below $m_t = 172.5~\GeV$  and NNLO masses slightly 
above.

\begin{figure}[htb]
\vspace{-1.5cm}
\centerline{\includegraphics[width=0.67\textwidth]{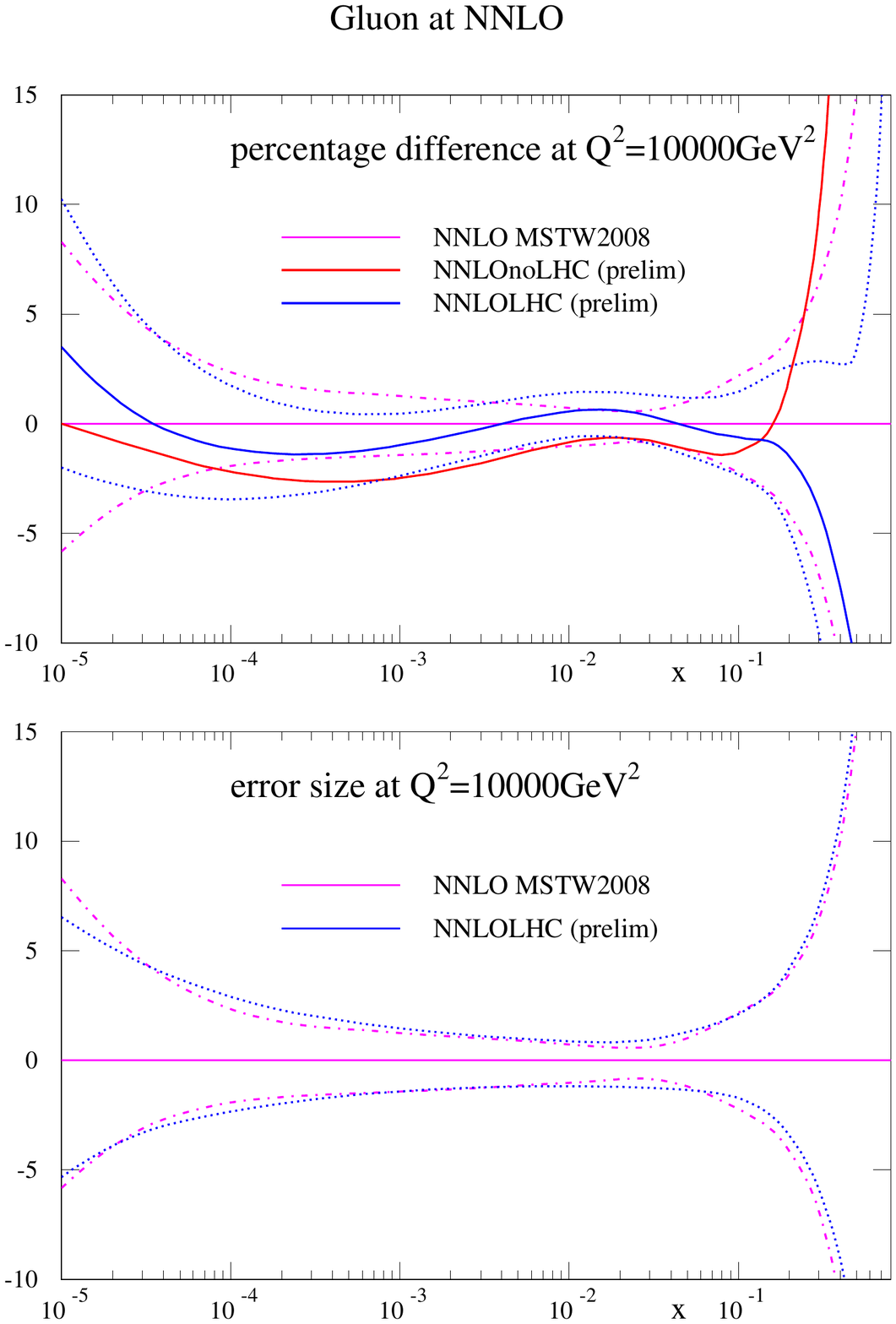}
\hspace{-2.6cm}\includegraphics[width=0.67\textwidth]{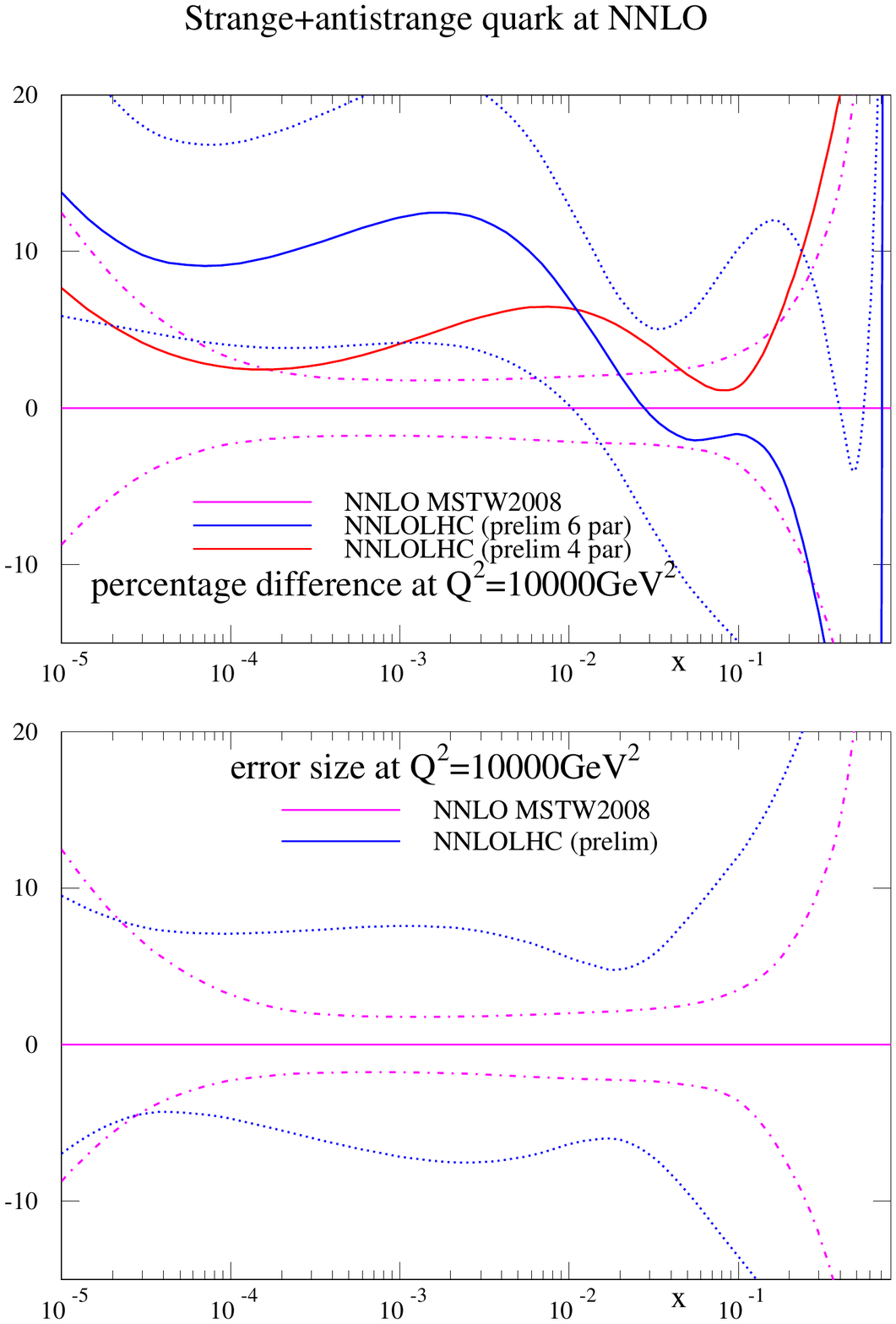}}
\vspace{-2.cm}
\caption{Comparison of the gluon and $s + \bar s$ distribution at NNLO. }
\vspace{-0.5cm}
\label{Fig3} 
\end{figure}

At NLO we also include CMS inclusive jet data data \cite{Chatrchyan:2013tia}
together with ATLAS 7~\TeV \cite{Aad:2011fc} and 2.76~\TeV data 
\cite{Aad:2013lpa}. 
The ATLAS fit quality is $\chi^2/N_{pts}=112/114$ and for CMS is  
$\chi^2/N_{pts}=186/133$ before the data are included 
directly -- at least as good as most other PDF sets.
(This does not include the further breakdown of one correlated uncertainty
into five recommended in \cite{CMS:2013yua}, which lowers the $\chi^2$
to about 1 per point.)  The simultaneous fit of 
CMS data together with ATLAS 7~\TeV + 2.76~\TeV data
leads to a reasonable improvement for CMS data, but only a tiny amount for 
ATLAS data. The inclusive jet data from the
experiments seem extremely compatible. 
Including all the LHC data at NLO leads to $\alpha_S(M_Z^2)=0.1193$,
close to the MSTW2008 value.
Whether to include inclusive jet data at NNLO is not clear. Previous
analyses have used the approximate threshold corrections 
\cite{Kidonakis:2000gi} which give about a 10-20$\%$ correction.
For LHC data the corrections are very similar for fairly high $x$ values 
such as those probed at the Tevatron, as illustrated in Fig. 50 of
\cite{Watt:2013oha}, but they blow up when low $x$ is
probed at the LHC, i.e. far from threshold. Moreover, the initial 
threshold calculation does not account for the variation with 
different jet radius $R$. A recent improved calculation \cite{deFlorian:2013qia}
has built in $R$ dependence and shows that while the $R$-dependence is large 
at NLO there is little extra $R$ variation at NNLO. However, the 
calculations still have problems at low $p_T$. 
The enormous project of the full 
NNLO calculation \cite{Ridder:2013mf,Currie:2013dwa} is nearing 
completion, and 
gives some indications of the full form of the correction, with correspondence 
to the approximate threshold correction in the appropriate region.
Hence, as default at NNLO we still fit 
Tevatron jet data, which seems safe since these are always relatively near to 
threshold. We do not include the LHC jet data in the standard fit.  
However we also try putting in ``smaller'' and ``larger'' approximate NNLO 
$K$-factors for the LHC data, i.e. with corrections of about $\sim 10\%$
and $\sim 20\%$ respectively at $p_T=100~\GeV$. The prediction for LHC data 
is good for the PDFs where the data are not included in the fit. 
The fit quality is a small amount worse 
than at NLO, and deteriorates a little with the larger $K$-factor.
At NNLO the extracted $\alpha_S(M_Z^2)=0.1162$, but with a larger
uncertainty in the upwards direction. 
When the LHC jet data are included in a fit the quality improves by a 
few units in $\chi^2$, mainly for CMS data, and both PDFs and 
$\alpha_S(M_Z^2)$ change by amounts very much smaller than their uncertainty. 
For most of the PDFs the change compared to MSTW2008 at NNLO 
is very similar to that at NLO. 
There are some differences for the gluon and strange quark,
and the central values obtained in our updated fit 
are shown at NNLO as a ratio to MSTW2008 in Fig. \ref{Fig3} . 
The uncertainty on the NNLO gluon is perhaps slightly larger at
high $x$ due to the omission of LHC jet data. The strange quark increases 
slightly more at NNLO than it does at NLO, though the details depend on 
the number of free parameters in the strange quark distribution, where 
6 free parameters gives some unusual features, so we reduce to four. 

These NLO and NNLO PDFs are not final, but we expect little change in an
updated set soon to be released.

\vspace{-0.4cm}

\section*{Acknowledgements}

\vspace{-0.2cm}

We would like to thank W. J. Stirling  and G. Watt for
numerous discussions on PDFs. This work is
supported partly by the London Centre for Terauniverse Studies (LCTS),
using funding from the European Research Council via the Advanced 
Investigator Grant 267352. RST would also like to thank the IPPP, Durham, for 
the award of a Research Associateship. We would like to thank the 
Science and Technology Facilities Council (STFC) for support.

\vspace{-0.4cm}

%\section{...}

\end{document}